\documentclass[12pt]{iopart}
\usepackage{iopams}  
\usepackage{graphicx}
\begin{document}

% TITLE PAGE {{{1
\title[Efficient polar convolution based on the dFBT]
{Efficient polar convolution based on the discrete 
Fourier-Bessel transform for application in 
computational biophotonics}

\author{O Melchert, M Wollweber and B Roth}

\address{Hanover Centre for Optical Technologies (HOT), 
Leibniz Universit\"at Hannover, 
Nienburger Str.\,17, 
D-30167 Hannover, Germany}

\ead{oliver.melchert@hot.uni-hannover.de}

\begin{abstract}
We discuss efficient algorithms for the accurate forward and reverse evaluation
of the discrete Fourier-Bessel transform (dFBT) as numerical tools to assist in
the $2$D polar convolution of two radially symmetric functions, relevant, e.g.,
to applications in computational biophotonics.
  In our survey of the numerical procedure we account for the circumstance
that the objective function might result from a more complex measurement
process and is, in the worst case, known on a finite sequence of 
coordinate values, only.
  We contrast the performance of the resulting algorithms with a procedure
based on a straight forward numerical quadrature of the underlying integral
transform and asses its efficienty for two benchmark Fourier-Bessel pairs. 
  An application to the problem of finite-size beam-shape convolution in polar
coordinates, relevant in the context of tissue optics and optoacoustics, is
used to illustrate the versatility and computational efficiency of the
numerical procedure.
\end{abstract}

% PACS numbers
% 02.70.-c -- Computational techniques in mathematical methods in physics
% 02.30.Gp -- Mathematical methods in physics. Special functions
% 02.30.Mv -- Mathematical methods in physics. Approximations and expansions
% 87.64.Aa -- Spectroscopic and microscopic techniques in biophysics and 
%                 medical physicsi. Computer simulation

% Keywords
\vspace{2pc}

% Submitted to journal title message
%\submitto{\MSMSE}

% Separate title page is required
%\maketitle

% TITLE PAGE 1}}}

% INTRODUCTION {{{1
\section{Introduction}

% Topic of the article:
% (1) Fourier Bessel intro
The Fourier-Bessel transform (FBT; also referred to as ``$0$th order Hankel
transform'') represents a mathematical tool that appears in numerous
computational approaches in science and engineering.  Among those are, e.g.,
applications in atomic scattering \cite{Talman:1978}, electron microscopy
\cite{FiskJohnson:1987}, and beam propagation through axially symmetric systems
\cite{GuizarSicairos:2004}.
The underlying theory and the operational rules for use with the FBT and, more
generally, the $n$th order Hankel transform are thoroughly discussed in
Ref.~\cite{Baddour:2015} where also an extensive review of the scientific
literature can be found.

% (2) Polar convolution intro
In addition to the above applications, the FBT allows for the convolution of
two radially symmetric functions in polar coordinates \cite{Baddour:2009}, a
computational tool in its own right.  This is viable since the general $2$D
convolution of two functions can be expressed in terms of their respective
Fourier series expansion, exhibiting a nontrivial relation to the $n$th order
(reverse) Hankel transform. 
However, if both functions are restricted to be radially symmetric, their
respective Fourier series expansions are nonzero for the term $n=0$ only, and,
consequently, their convolution can be shown to relate to a FBT, see, e.g., 
Ref.~\cite{Baddour:2009} which elaborates  on the minutiae of this issue.  

% Motivation 
In the presented article, we aim to draw some more attention to an efficient 
algorithm for the accurate evaluation of the discrete Fourier-Bessel transform 
(dFBT) due to Fisk-Johnson \cite{FiskJohnson:1987}. Albeit the latter reference
introduced the dFBT algorithm, an in-depth discussion of the discretization 
scheme for the forward and reverse transformation are provided by 
Ref.~\cite{Baddour:2015}.
Our motivation to study the Fisk-Johnson dFBT procedure is based on its
efficiency for the purpose of polar convolution. As discussed in 
the seminal article \cite{FiskJohnson:1987}, the algorithmic procedure might 
lead to a significant reduction in computation time, if, subsequent to a
dFBT a follow up back transformation is required. 
Here, we present a particular application in computational biophotonics where 
this comes in handy. 
More precisely, we consider a problem in tissue optics where the task is to
convolve the Green's function response of a (possibly) multilayered tissue 
with a custom irradiation source profile to yield the response to a laser-beam 
of finite diameter.
Therein, the Green's function response is obtained from computer simulations 
involving an infinitely thin ``pencil'' laser-beam \cite{MCML:1995}, thus 
resulting from a complex measurement process that yields the obective 
function on a finite sequence of equidistant sample points.

% Outline
The article is organized as follows: in section \ref{sec:dFBT} we resume the
forward and reverse dFBT procedures, paving the way for an efficient polar
convolution algorithm, followed by an assessment of their accuracy and
perfomance for two benchmark Fourier-Bessel transform pairs in section
\ref{sec:examples}. In section \ref{sec:polarConv} we then elaborate on the
problem of postprocessing a Green's function material response to conform to a
spatially extended photon beam in computational biophotonics.  Finally, in
section \ref{sec:fin} we summarize and conclude on the presented study.
% 1}}}

% DISCRETE FOURIER-BESSEL TRANSFORM 1{{{
\section{Polar convolution in terms of discrete Fourier-Bessel transforms (dFBTs)}
\label{sec:dFBT}

Here we consider a discrete approximation to the Fourier-Bessel transform
$F_0(\rho)$ of a function $f(x)$ of a real variable $x\geq 0$, for which 
$\int_0^\infty f(x) x^{1/2}~{\rm d}x$ is required to be absolutely 
convergent, defined by 
\cite{Baddour:2009,Baddour:2015} 
\begin{equation}
F_0(\rho) = \int_0^\infty f(x) J_0(x\rho) x~{\rm{d}}x. \label{eq:FWD}
\end{equation}
Due to self-reciprocality, its reverse transform reads
\begin{equation}
f(x) = \int_0^\infty F_0(\rho) J_0(x\rho) \rho~{\rm{d}}\rho. \label{eq:BCKWD}
\end{equation}
Therein $J_0(\cdot)$ signifies the $0$th order Bessel function and, together,
$f$ and $F_0$ comprise a Fourier-Bessel transform pair.  Following
Ref.~\cite{FiskJohnson:1987}, the dFBT is based on two reasonable assumptions:
(A1) one can give a truncation threshold $T$ above which the objective function
vanishes, and, (A2) the Fourier-Bessel series of the objective function might
be truncated after $N$ terms.  From an applied point of view and so as to
yield a finite computational procedure, both assumptions are inevitable
and might be satisfied by reasonably large values of $T$ and $N$.
Subsequently, we distinguish the forward transform for continuous objective
functions as well as for objective functions known at a finite sequence sample
points and allude to their universal backward transformation.

\paragraph{Forward transform for continuous objective functions -}

For given values of $T$ and $N$, let $\{j_m\}_{m=1}^N$ denote the sequence of the
first $N$ zeros of $J_0$ in ascending order. Then, the forward dFBT for a
continuous objective function, involving the zeros of the Bessel function, 
derived in Ref.~\cite{FiskJohnson:1987}, reads
\begin{equation}
F_0(j_m/T) = \frac{2 T^2}{j_N^2} \sum_{k=1}^{N-1} \frac{J_0(j_k j_m/ j_N)}{J_1^2(j_k)} f(j_k T/j_N), \label{eq:dFBTC}
\end{equation}
where $J_1(\cdot)$ refers to the first order Bessel function.  The above
approximation to Eq.\,(\ref{eq:FWD}) is feasible, since, given a continuous
objective function, the function values at $\{x_k T\}_{k=1}^{N-1}$ with
$x_k=j_k/j_N$ can be computed in a straight forward manner. As a result one
obtains the Fourier-Bessel transform of $f(x)$ at the discrete sequence
$\{j_m/T\}_{m=1}^N$ of scaled Bessel zeros.  Note that the above algorithm
terminates in time $O(N^2)$.

\paragraph{Forward transform for discrete objective functions -}

If the objective function is known for a finite sequence $\{x_k T\}_{k=1}^M$ of
sample points that do not meet the requirement of $x_k=j_k/j_N$ in
Eq.~(\ref{eq:dFBTC}) above, we might nevertheless proceed by computing its
Fourier-Bessel expansion coefficients to obtain its transform at the same set
$\{j_m/T\}_{m=1}^N$ of sample points as
\begin{equation}
F_0(j_m/T) = T^2 \int_0^1 x f(x T) J_0(j_m x)~{\rm{d}}x \label{eq:dFBTD}
\end{equation}
provided that the number of sample points $M$ is large enough. In our numerical
experiments we used a trapezoidal rule to approximate the integral in
Eq.~(\ref{eq:dFBTD}).
Note that under the reasonable assumption $M\gg N$, the above algorithm 
terminates in time $O(NM)$.

\paragraph{Universal backward transform -}

If, subsequent to one of the transforms given by Eqs.~(\ref{eq:dFBTC}) and 
(\ref{eq:dFBTD}), an immediate back-transformation is required, arbitrary 
function values $f(x T)$ for the parameters $T$ and $N$ can be computed by 
using the sequence $\{F_0(j_m/T)\}_{m=1}^N$ of dFBT samples
according to \cite{FiskJohnson:1987}
\begin{equation}
f(x T) = \frac{2}{T^2} \sum_{m=1}^{N-1} \frac{F_0(j_m/T)}{J_1^2(j_m)} J_0(j_m x), \qquad{0\leq x \leq 1}. \label{eq:dFBTRev}
\end{equation}
Note that due to (A1) one has $f(x T) = 0$ for $x>1$.
Further, note that the above reverse algorithm terminates in time $O(N)$ for
a given value of $x$.

\paragraph{Polar convolution using the dFBT -}

As pointed out earlier, from a point of view of computational complexity, the 
Fisk-Johnson procedure is particularly efficient if a dFBT, resulting
in the sequence of transform estimates $\{F_0(j_m/T)\}_{m=1}^{N}$, is followed 
by a reverse transform based on the summation of $F_0$ at the exact same 
sequence of sample points along the transformed domain. 
Now, considering two radially symmetric functions it is possible to take
advantage of the above procedure in order to derive an efficient algorithm for
their polar convolution. Let $f(r)$ and $g(r)$ be two such radially symmetric
functions. Then, their $2$D (polar) convolution $h(r)$, again a function with
radial symmetry, can be computed via \cite{Baddour:2009} 
\begin{equation}
h(r) = {\mathsf{polConv}}[f,g](r) = 2 \pi \int_{0}^\infty 
F_0(\rho) G_0(\rho) J_0(\rho r)\rho~{\rm{d}}\rho, \label{eq:polConv}
\end{equation}
wherein $F_0(\rho)$ and $G_0(\rho)$ signify the Fourier-Bessel transforms of
$f(r)$ and $g(r)$, respectively \cite{comment:polConv}. A Fisk-Johnson 
approximation ${\mathsf{polConv}}[f,g](r;T,N)$ of the polar convolution can thus be 
formulated as a three step procedure:
(i) set the threshold parameters $T$ and $N$ of the Fisk-Johnson procedure,
(ii) compute both dFBTs $F_0(\rho_m)$ and $G_0(\rho_m)$
at the same sequence $\{\rho_m = j_m/T\}_{m=1}^N$ of samples along the 
transformed domain using either Eq.~(\ref{eq:dFBTC}) or (\ref{eq:dFBTD}), and,
(iii) compute the pointwise products $H_0(\rho_m)=2\pi F_0(\rho_m)G_0(\rho_m)$ 
followed by a reverse transformation via Eq.~(\ref{eq:dFBTRev}) to yield
$h(xT)$ for $0\leq x \leq 1$.
The resulting Fisk-Johnson polar convolution is thus no more expensive than
$O(NM)$ if the number of grid points $x_i$ at which $h(x_iT)$ is sampled  
is of order $O(M)$.
% 1}}}

% BENCHMARKING VIA KNOWN FB PAIRS 1{{{
\section{Benchmarking via known Fourier-Bessel pairs}
\label{sec:examples}

\begin{figure}[t!]
\centerline{\includegraphics[width=0.75\linewidth]{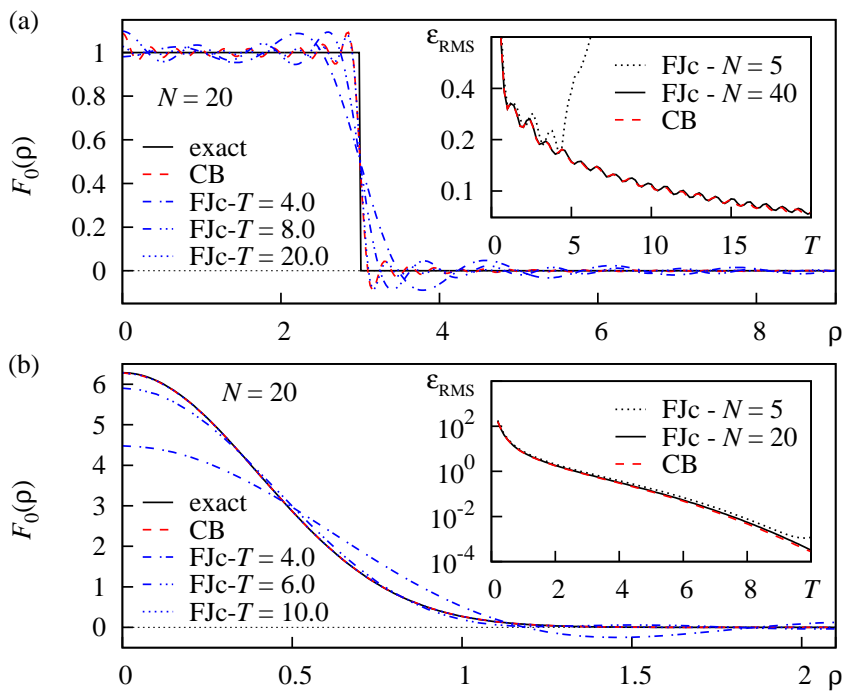} } 
\caption{(Color online)  Discrete Fourier-Bessel transform (dFBT) for two 
benchmark transform pairs and different transform parameters $T$ and $N$,
following the method of Fisk-Johnson for a continuous objective function
discussed in section \ref{sec:examples}.
(a) Transform of the {\rm jinc}-function discussed in the text. The main
plot shows a sequence of extrapolated function values of the transform for
differenct choices ot the truncation parameter $T$ at $N=20$ (blue dash-dotted
curves; labeled FJc). The result obtained by a straight forward numerical
integration of the integral transform is shown as red dashed curve (labeled
CB).  The inset illustrates the root-mean squared error $\epsilon_{\rm RMS}(T)$
for two choices of $N$. 
(b) Transform of the Gaussian function discussed in the text. The main
plot shows the extrapolated function values for different choices of $T$ at
$N=20$ and the inset illustrates $\epsilon_{\rm RMS}$ for two values of $N$.}
\label{fig:examples}
\end{figure}

So as to compare the Fisk-Johnson dFBT of an objective function, represented by
the sequence of $N-1$ values $\{F_0(j_m/T)\}_{m=1}^{N-1}$, to the exact
transform, we need to agree upon a representative sequence of coordinate values
of the transformed grid at which to evaluate both. Here we proceed as follows:
we consider a further ``benchmark'' method wich was previously assessed, and,
albeit being computationally rather inefficient, reported to be quite precise
\cite{Cree:1993}. We refer to this reference method as the ``Cree-Bones''
(CB) algorithm, implemented as a numerical integration of the integral transform
Eq.~(\ref{eq:FWD}) using a trapezoidal rule and grid partitioning as reported
in Ref.~\cite{Cree:1993}. For comparison, if the objective function is
available at $M$ grid points, the CB algorithm terminates in time $O(M^2)$.
Subsequently, considering a Fourier-Bessel transform pair, we
compute the dFBT using the Fisk-Johnson and Cree-Bones procedures. The latter
yields a sequence of coordinates $\{\rho_i\}_{i=0}^{M-1}$ at which we 
evaluate the exact transform and for which we extrapolate the Fisk-Johnson 
dFBT using \cite{FiskJohnson:1987}
\begin{equation}
F_0(\rho_i) = 2 \sum_{m=1}^{N-1} \frac{j_m J_0(\rho_i T)}{J_1(j_m)(j_m^2-\rho_i^2 T^2)} F_0(j_m/T). \label{eq:extrapolation}
\end{equation}
As illustrated in Fig.~\ref{fig:examples}, this not only allows to visually
assess the performance of the Fisk-Johnson dFBT procedure for different choices
of the truncation parameters $T$ and $N$, but also allows to quantify the 
deviation from the exact transform in tems of the relative root-mean-squared 
error
\begin{equation}
\epsilon_{\rm RMS} = \left( \frac{\sum_i [F_0^{\rm{exact}}(\rho_i) - F_0^{\rm dFBT}(\rho_i)]^2}{\sum_i [F_0^{\rm dFBT}(\rho_i)]^2} \right)^{-1/2}. \label{eq:rms}
\end{equation}

%%% BEGIN: FIGURE %%%%%%%%%%%%%%%%%%%%%%%%%%%%%%%%%%%%%%%%%%%%%%%%%%%%%%%%%%%%%
%
% EXAMPLARY APPLICATION TO TWO FOURIER-BESSEL PAIRS
%
%%%%%%%%%%%%%%%%%%%%%%%%%%%%%%%%%%%%%%%%%%%%%%%%%%%%%%%%%%%%%%%%%%%%%%%%%%%%%%%
%%% END: FIGURE %%%%%%%%%%%%%%%%%%%%%%%%%%%%%%%%%%%%%%%%%%%%%%%%%%%%%%%%%%%%%%%

\paragraph{dFBT of a {\rm jinc} function -}
First we considered the Fourier-Bessel pair
\begin{equation}
f(r) = a_0^2 \,{\rm jinc}(a_0 r) \quad \Longleftrightarrow \quad 
F_0(\rho) = \theta(\rho-a_0),
\label{eq:FBP1}
\end{equation}
for $a_0=3$, wherein ${\rm jinc}(x)=J_1(x)/x$ and $\theta(\cdot)$ signifies the
Heavyside step function.  While here, the Fisk-Johnson algorithm exploits the
possibility to compute $f(r)$ at precisely those sample points required by
Eq.~(\ref{eq:dFBTC}), the Cree-Bones algorithm used an equispaced grid
$\{r_i\}_{i=0}^{M-1}$ with, $r_i=r_0+i\Delta$ and
$\Delta=(r_{M-1}-r_{0})/(M-1)$ where $r_0 = 0.01$, $r_{M-1}=20$, $M=1000$.  As
usual in Fourier-type function approximation, due to the discontinuous nature
of $\theta$, we expect this kind of benchmark transform pair to represent a
difficult test for any kind of dFBT.  Bearing this in mind, the considered
transform pair might be regarded as a worst-case use case that might arise in
computational biophotonics since a commonly employed irradiation source profile
(ISP), referred to as ``top-hat'' ISP, exhibits the shape of $\theta$
\cite{Paltauf:2000,CONV:1997}.  Consequently, any convolution using such an ISP
involves a revese dFBT of the above form.

In Fig.~\ref{fig:examples}(a) we show the result of applying the dFBT to the
above {\rm jinc}-function. In the main plot of Fig.~\ref{fig:examples}(a) we
explore the effect a finite truncation threshold $T$ has on the transformed
function for the summation threshold $N=20$. Note that for small values of $T$
the Fisk-Johnson dFBT deviates significantly from the exact transform (solid
black line). This is due to the assumption that above $T$ the objective
function vanishes and, hence, structural details of the {\rm jinc}-function
bejond that threshold are ignored in the transformation process.  As the value
of $T$ grows larger, the dFBT approximation at sufficiently large $N$ gets
increasingly better as shown by the overall decrease of the RMS error in the
inset. While, at given $T$, a too small value of $N$ leads to a huge RMS error,
reflecting that the truncated Fourier-Bessel series has not converged as in the
case of $N=5$, the accuracy of the Fisk-Johnson transform at $N=40$ is similar
to that of the (computationally more expensive) Cree-Bones transform.
To support intuition, further computer experiments indicate that, e.g., at
$T=10$ there exists a narrow threshold range $N=6-12$ within which the RMS
error decreases by one order of magnitude (not shown; see discussion below), and
where $\epsilon_{\rm RMS}(N>12)\approx 0.12$ (cf.\ inset of
Fig.~\ref{fig:examples}(a)).  For completeness, note that for $T=20$ and
$N=20$, the Fisk-Johnson and Cree-Bones dFBT agree well as illustrated in the
main plot of Fig.~\ref{fig:examples}(a). Both feature Gibbs ringing artifacts
that might be expected for this kind of transform pair.

\paragraph{dFBT of a Gaussian function -}
Next, we consider the dFBT transform of a Gaussian function
\begin{equation}
f(r) = \exp\{-r^2/(4\pi)\} \quad \Longleftrightarrow \quad F_0(\rho)=2 \pi \exp\{-\pi r^2\}. \label{eq:FBP2}
\end{equation}
For the numerical experiments using the Cree-Bones algorithm we again used an
equispaced grid $\{r_i\}_{i=0}^{M-1}$ with, $r_i=r_0+i\Delta$ and
$\Delta=(r_{M-1}-r_{0})/(M-1)$ where $r_0 = 0.01$, $r_{M-1}=10$, $M=1000$. For this
kind of smooth benchmark transform pair we expect the accuracy of the tranform
to be even better than in the previous case.  This type of objective function
might be regarded as a best-case use case that might arise in computational
biophotonics since another commonly employed ISP has the shape of a simple
Gaussian function \cite{Paltauf:2000,CONV:1997}. 

As evident from Fig.~\ref{fig:examples}(b), similar to the previous example, if
the truncation threshold $T$ is chosen too small, the transfrom deviates from
the exact result since vital parts of the objective function beyond $T$ are
ignored. To support intuition, note that $f(r)$ drops to its $1/e$-height at
$T=2 \pi^{1/2}\approx 3.5$, explaining the deviation of the $T=4$ dFBT
approximation to the exact result.
However, a visual inspection of the approximation at $T=10$, where one finds 
\mbox{$f(0)/f(10)\approx 2.8\cdot 10^3$}, indicates that it fits the asymptotic
result quite well. This ``chi-by-eye'' result is supported by the relative RMS
error illustrated in the inset. Even at small values of the summation trunction
parameter $N$, the accuracy of the Fisk-Johnson dFBT improves noticably as
$T\to 10$ and approaches the approximation error of the CB transform at a given
value of $T$ rapidly as $N$ is adjusted to higher values. 

%%% BEGIN: FIGURE %%%%%%%%%%%%%%%%%%%%%%%%%%%%%%%%%%%%%%%%%%%%%%%%%%%%%%%%%%%%%
%
% EXAMPLARY FISK-JOHNSON REVERSE TRANSFORM FOR THE JINC FUNCTION
%
%%%%%%%%%%%%%%%%%%%%%%%%%%%%%%%%%%%%%%%%%%%%%%%%%%%%%%%%%%%%%%%%%%%%%%%%%%%%%%%
\begin{figure}[t!]
\centerline{\includegraphics[width=0.75\linewidth]{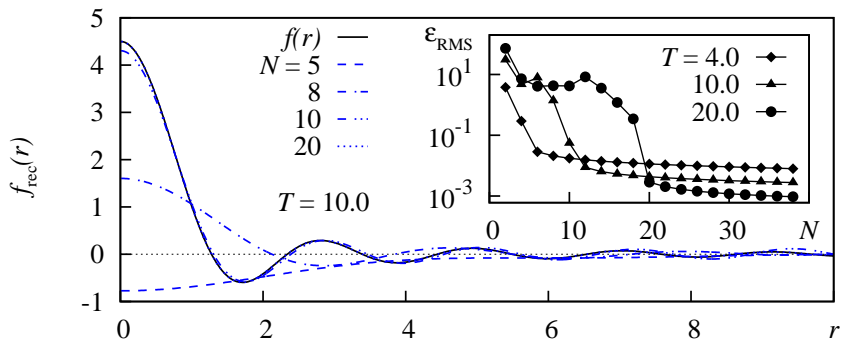} } 
\caption{(Color online) Reconstruction $f_{\rm rec}$ of the initial function
under a reverse dFBT. The main plot shows the original objective function
(solid black curve) and the reconstructed functions at $T=10.0$ for different
values of $N$. The inset illustrates the relative RMS error (see
Eq.~\ref{eq:rms}) for three choices of the truncation threshold $T$ for $N=2$
through $40$.}
\label{fig:reverseTrafo}
\end{figure}
%%% END: FIGURE %%%%%%%%%%%%%%%%%%%%%%%%%%%%%%%%%%%%%%%%%%%%%%%%%%%%%%%%%%%%%%%

\paragraph{Reconstruction of the objective function -}
Next, we assess the accuracy of a reconstruction of the objective function
under a reverse dFBT implemented according to Eq.~(\ref{eq:dFBTRev}).
Therefore, we first compute the dFBT approximation to the {\rm jinc}-objective
function using the sequence of grid samples required by Eq.~(\ref{eq:dFBTC}),
where we considered the truncation threshold $T=10.0$ and different values of
$N$. The results of a subsequent reverse transformation, computed for a
sufficiently sampling density of $x$ via the Fisk-Johnson procedure are
summarized in Fig.~\ref{fig:reverseTrafo}.  As evident from the main plot of
the figure, the reconstruction of the objective function seems to be quite
accurate once the summation truncation parameter exceeds $N=20$. This finding
can be put on a more quantitative basis by means of the relative RMS error,
reported in the inset of Fig.~\ref{fig:reverseTrafo}. We find that at $T=10.0$
there exists a narrow threshold range $N=6-12$ within which the RMS error
decreases by almost three orders of magnitude from $\epsilon_{\rm
RMS}(N=6)\approx 7.7$ to $\epsilon_{\rm RMS}(N=12)\approx 0.007$. For higher
(smaller) values of $T$, this threshold range can be seen to shift towards
higher (smaller) values of $N$.  This is intuitive since at larger values of
$T$ more sample points of the transformed domain are necessary to capture the
structural details of the underlying function appropriately, thus affecting the
convergence of the truncated sums used to approximate the Fourier-Bessel
integral transform. 

%%% BEGIN: FIGURE %%%%%%%%%%%%%%%%%%%%%%%%%%%%%%%%%%%%%%%%%%%%%%%%%%%%%%%%%%%%%
%
% EXAMPLARY APPLICATION TO TWO POLAR CONVOLUTION 
%
%%%%%%%%%%%%%%%%%%%%%%%%%%%%%%%%%%%%%%%%%%%%%%%%%%%%%%%%%%%%%%%%%%%%%%%%%%%%%%%
\begin{figure}[t!]
\centerline{\includegraphics[width=0.75\linewidth]{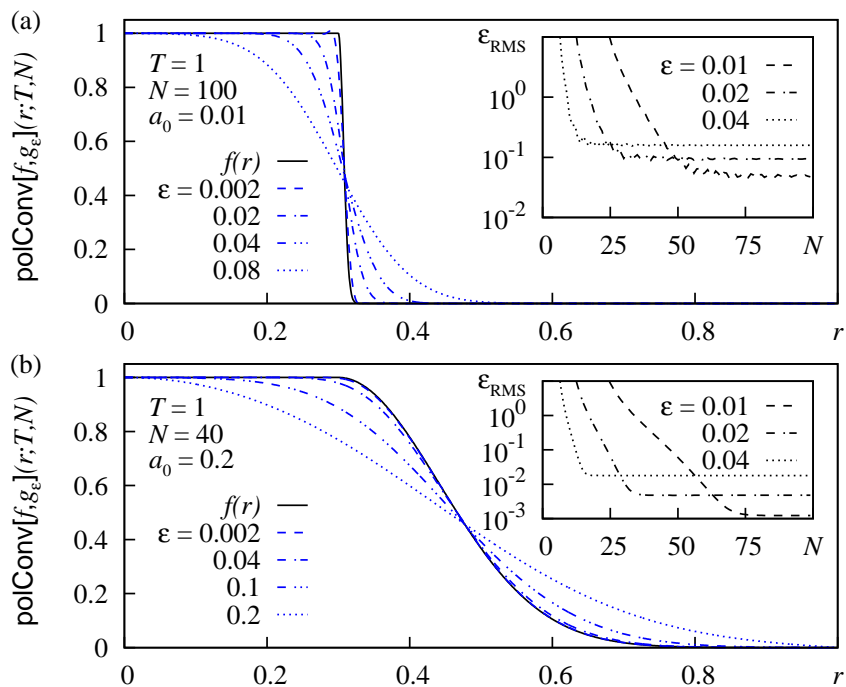} } 
\caption{(Color online) Exemplary $2$D polar convolution using the Fisk-Johnson
procedure detailed in section~\ref{sec:dFBT}. The figure illustrates the 
convolution $h(r) = {\mathsf{polConf}}[f,g](r;T,N)$ of a flat-top profile $f(r)$ with 
a Gaussian approximation $g_\epsilon(r)$ to the delta-distribution, see 
section~\ref{sec:examples}.
In the limit $\epsilon \to 0$ we expect to find $h(r)\to f(r)$.
(a) Convolution using the ``steep'' flat-top parameters $r_0=0.3$, $a_0=0.01$ and truncation
thresholds $T=1$, $N=100$ for different values of $\epsilon$. The inset shows
the relative RMS errors for the approximation of $f(r)$ by $h(r)$ as function 
of the summation truncation parameter $N$. 
(b) same as (a) for ``smooth'' flat top parameters $r_0=0.3$, $a_0=0.2$ and
summation truncation parameter $N=40$. 
}
\label{fig:polConv}
\end{figure}
%%% END: FIGURE %%%%%%%%%%%%%%%%%%%%%%%%%%%%%%%%%%%%%%%%%%%%%%%%%%%%%%%%%%%%%%%
\paragraph{Exemplary polar convolution -}
Finally, we test the performance of the dFBT for the purpose of $2$D polar 
convolution. Therefore, we consider the two functions
\numparts
\begin{eqnarray}
f(r) = 
\cases{1 &for $r \le r_0$\\
\exp\{-(r-r_0)^2/a_0^2\} &for $r>r_0$\\},\label{eq:flatTop}\\
g_\epsilon(r) =(2\pi \epsilon^2)^{-1} \exp\{-r^2/(2\epsilon^2))\}\label{eq:delt}.
\end{eqnarray}
\endnumparts
and follow the procedural description detailed in section~\ref{sec:dFBT}
to compute ${\mathsf{polConv}}[f,g_\epsilon]$. 
Note that Eq.~(\ref{eq:flatTop}) represents a ``flat-top'' ISP, i.e.\ a top-hat
function with a smooth roll-off, consistent with actual beam profiles observed in
laboratory experiments, see Refs.~\cite{Paltauf:1997, Paltauf:1998,
Paltauf:2000, DAlessandro:2012, Blumenroether:2016} that report on 
flat-top ISPs with parameter ratio in the range $r_0/a_0=1-10$.  
Further, Eq.~(\ref{eq:delt}) signifies a Gaussian approximation to a $2$D
delta-function, attained in the limit $\epsilon \to 0$. Hence, we expect to 
find \mbox{$\lim_{\epsilon \to 0} {\mathsf{polConv}}[f,g_\epsilon](r)= f(r)$}.
In this question, Fig.~\ref{fig:polConv} illustrates the accuracy of the
Fisk-Johnson convolution procedure for a ``steep'' example with $r_0/a_0=300$
and a ``smooth'' example with $r_0/a_0=1.5$, see Figs.~\ref{fig:polConv}(a) and
(b), respectively. As evident from the scaling behavior of the associated 
RMS error between $f(r)$ and $h(r)$ (shown in the inset of the subfigures), the 
accuracy of the approximation at fixed $T=1.0$ and given $\epsilon$ 
increases as the summation truncation parameter $N$ increases, saturating
at a characteristic limiting value $N_\epsilon$. As $\epsilon$ decreases, i.e.\
the closer $g_\epsilon(r)$ approximates a delta-function, the approximation
error of $h(r)$ also decreases. Bearing in mind the above results for the
forward and reverse dFBT it does not come as a surprise that the polar
convolution of a ``smooth'' objective function with a delta-function is more
accurate than that of a ``steep'' objective function.
% 1}}}

% APPLICATION TO BEAM SHAPE CONVOLUTION {{{1
\section{Application to beam-shape convolution in polar coordinates}
\label{sec:polarConv}

% (0) Consider layered homogeneous absorbing/scattering media.
An application of the efficient Fisk-Johnson polar convolution algorithm to a
particular problem in computational biophotonics is illustrated in the
remainder. It provides a solution to the issue of computing the material
response to custom radially symmetric laser beams of finite extend for layered
homogeneous media, given the corresponding Green's function response of the
medium. 
To illustrate the computational procedure we considered the simple but
paradigmatic case of a semi-infinite medium with a refractive-index-mismatched
boundary. For the optical parameters we used the relative refractive indices
$n=1.0$ (for the ambient medium) and $n=1.37$ as well as the absorption
coefficient $\mu_{\rm a}=0.1\,{\rm cm^{-1}}$, scattering coefficient $\mu_{\rm
s}=10.0\,{\rm cm^{-1}}$ and values of the anisotropy parameter
$g\in[0.1,0.95]$.  

%%% BEGIN: TABLE %%%%%%%%%%%%%%%%%%%%%%%%%%%%%%%%%%%%%%%%%%%%%%%%%%%%%%%%%%%%%%
%
% SIMULATION PARAMETERS AND CHARACTERISTIC LENGTH SCALES 
%
%%%%%%%%%%%%%%%%%%%%%%%%%%%%%%%%%%%%%%%%%%%%%%%%%%%%%%%%%%%%%%%%%%%%%%%%%%%%%%%
\Table{\label{tab:simPars} 
Characteristic lengthscales \cite{Wilson:1990} for light transport in the
considered tissue setup and homogeneous grid parameter for discretization of
the source volume using {\tt MCML} \cite{MCML:1995}. From left to right:
anisotropy parameter $g$, mean free path (mfp) length $\ell_{\rm mfp}$,
transport mfp $\tilde{\ell}_{\rm mfp}$, penetration depth $d_{\rm p}$ and grid
parameters for the cylindrical 
sampling lattice.}
\br
&&&&\centre{3}{$z$-axis}& \centre{3}{$r$-axis} \\
\ns
&&&&\crule{3}&\crule{3}\\
$g$ & $\ell_{\rm mfp}$  & $\tilde{\ell}_{\rm mfp}$   & $d_{\rm p}$ & $N_z$ & $\Delta_z$ & $z_{\rm max}$ &   $N_r$ & $\Delta_r$ & $r_{\rm max}$   \\
 & (cm)  & (cm)  & (cm) & (bins) & (cm) & (cm) & (bins) & (cm) & (cm)   \\
\mr
0.10 & 0.099 & 0.110 & 0.605 & 363  & 0.005 & 1.815 & 1000 & 0.002 & 2.0   \\
0.70 & 0.099 & 0.323 & 1.037 & 622  & 0.005 & 3.11 & 1000 & 0.0033 & 3.3   \\
0.90 & 0.099 & 0.909 & 1.741 & 1044 & 0.005 & 5.22 & 1000 & 0.0053 & 5.4   \\
0.95 & 0.099 & 1.667 & 2.357 & 1414 & 0.005 & 7.07 & 1000 & 0.0073 & 7.3   \\
\br
\end{tabular}
\end{indented}
\end{table}
%%%%%%%%%%%%%%%%%%%%%%%%%%%%%%%%%%%%%%%%%%%%%%%%%%%%%%%%%%%%%%%%%%%%%%%%%%%%%%%

% (1) GreensFunc via MCML
\paragraph{Monte Carlo modelling of the Greens function response -}
For our numerical experiments we computed the Green's function $G$ of the
absorbed energy density for the above setup as the material response to an
infinitely thin ``pencil'' beam using the publicly available {\rm C} code {\tt
MCML} \cite{MCML:1995}. It solves the problem of steady-state light transport
in terms of a Monte Carlo approach to photon migration in layered media and
provides the accumulated observables on a homogeneous polar grid, i.e.\
$G\equiv G(r,z)$.  
For our numerical experiments we used the simulation parameters listed in
Tab.~\ref{tab:simPars}. In setting up the discretized source volume we made
sure the maximal $z$-depth $z_{\rm max}$ and $r$-range $r_{\rm max}$ exceed the
penetration depth $d_{\rm p}$ of photons within the medium by a factor of three
at least. Note that for extended beam profiles and not too close to the
material surface, $d_{\rm p}$ refers to the intrinsic length-scale after which
the fluence-rate along the beam-axis reduces to its $1/e$-value
\cite{MCML:1995,Wilson:1990}.  For completeness, one might perform the
numerical experiments as well by one of {\tt MCML}s descendants designed for
layered homogeneous media, as, e.g., {\tt GPU-MCML} \cite{Alerstam:2010}.

\paragraph{Material response to laser beams with finite extend -}
% (2) Given GreensFunc -> distribution of absorbed energy density W.
In order to obtain the desired material response $W(r,z)$ to an extended
radially symmetric laser beam, the Green's function $G(r,z)$ needs to be
convolved using an appropriate transverse ISP $f(r)$.  
% (3) Issues with CONV
In principle this can be done using the publicly available {\tt C} 
code {\tt CONV} \cite{CONV:1997}, that implements a top-hat and a Gaussian ISP.  
However, note that since {\tt CONV} features only these two ISPs it is of 
rather limited use. Albeit allowing for a highly efficient direct convolution
involving the solution of $1$D integrals only, both beam profiles are not
consistent with actual profiles observed in laboratory experiments, see
Refs.~\cite{Paltauf:1997, Paltauf:1998, Paltauf:2000, DAlessandro:2012,
Blumenroether:2016}.  
% (4) Remendy: general purpose routive needed 
Further, on a more general basis, a computationally efficient and more
versatile solution procedure that allows for convolution with custom ISPs seems
to be of value.

%%% BEGIN: FIGURE %%%%%%%%%%%%%%%%%%%%%%%%%%%%%%%%%%%%%%%%%%%%%%%%%%%%%%%%%%%%%
%
% BEAM SHAPE CONVOLUTION IN BIOPHOTONICS 
%
%%%%%%%%%%%%%%%%%%%%%%%%%%%%%%%%%%%%%%%%%%%%%%%%%%%%%%%%%%%%%%%%%%%%%%%%%%%%%%%
\begin{figure}[t!]
\centerline{\includegraphics[width=0.95\linewidth]{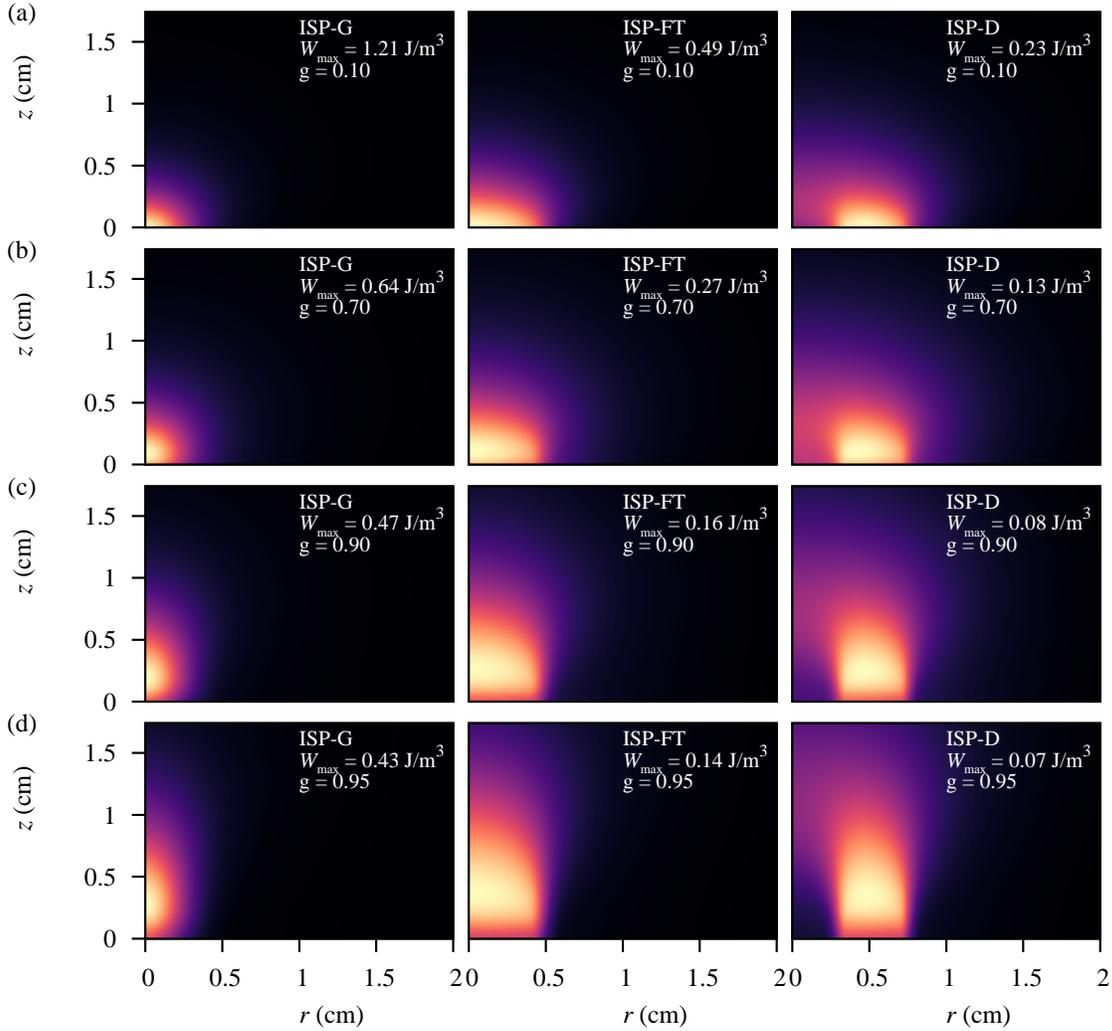} } 
\caption{(Color online) Illustration of the beam shape convolution procedure
for a Gaussian ISP (G; a special case of Eq.~(\ref{eq:donut}) with $r_0=0$ and
$r_1=0$) with parameter $a_1=0.25$, flat-Top ISP (FT; a special 
case of Eq.~(\ref{eq:donut}) with $r_0=0$) with parameters $r_1=0.4$ and $a_1=0.1$, 
and donut (D) ISP with parameters $r_0=0.25$, $r_1=0.6$ and $a_0=a_1=0.05$ 
considering four different values of the anisotropy parameter $g$.
(a) from left to right (ltr): G, FT and D ISP for $g=0.10$, 
(b) ltr: G, FT and D ISP for $g=0.70$, 
(c) ltr: G, FT and D ISP for $g=0.90$, 
(d) ltr: G, FT and D ISP for $g=0.95$. 
The maximal value of absorbed laser energy $W_{\rm max}~ ({\rm J/m^3})$
for each configuration, indicated by the brightest color, is listed within the
individual subfigures.
}
\label{fig:beamShapes}
\end{figure}
%%% END: FIGURE %%%%%%%%%%%%%%%%%%%%%%%%%%%%%%%%%%%%%%%%%%%%%%%%%%%%%%%%%%%%%%%

% (5) laser beams with finite extend
In this regard we follow a different approach by solving the 2D convolution
problem in terms of the Fourier-Bessel transform in polar coordinates
\cite{CONV:1997,Baddour:2009} 
\begin{equation}
W(r,z) = \mathsf{polConv}[f,G](r,z) = 2 \pi f_0 \int_0^\infty G_0(\rho,z) F_0(\rho) J_0(\rho r)\rho~{\rm d}\rho, \label{eq:polConvW}
\end{equation}
following the Fisk-Johnson discretization procedure detailed in section
\ref{sec:dFBT}. 
Therein, $f(r)$ signifies a custom ``donut'' ISP
\begin{equation}
f(r) = 
\cases{
\exp\{-(r-r_0)^2/a_0^2\} &for $r<r_0$\\
1 &for $r_0\le r \le r_1$\\
\exp\{-(r-r_1)^2/a_1^2\} &for $r>r_1$\\},\label{eq:donut}
\end{equation}
and $G$ stands for the laser absorption Green's function computed for an 
infinitely narrow laser beam, incident upon the material surface.
The respective dFBTs are given by $F_0$ and $G_0$.
In the above equation, $f_0$ allows to scale the beam intensity to achieve 
a total beam power $P$ via
\begin{eqnarray}
f_0 = P~\big[2 \pi \int_0^\infty\, r~f(r)~{\rm d}r\big]^{-1}.
\end{eqnarray}
Note that this yields a general purpose routine that allows for quite arbitrary
beam profiles, only required to obey the integrability conditions of a
Fourier-Bessel transform.
As a technicality, note that the dFBT $F_0$ of
the continuous ISP $f$, computed using the $O(M^2)$ algorithm
Eq.~(\ref{eq:dFBTC}), can be reused at each value of $z$.  In contrast to
the later function, since $G(r,z)$ is known at a finite number of sample
points only, its dFBT $G_0$ is obtained via the $O(NM)$ algorithm
Eq.~(\ref{eq:dFBTD}).

%%% BEGIN: FIGURE %%%%%%%%%%%%%%%%%%%%%%%%%%%%%%%%%%%%%%%%%%%%%%%%%%%%%%%%%%%%%
%
% Wrz SLICES ALONG FIXED z, r 
%
%%%%%%%%%%%%%%%%%%%%%%%%%%%%%%%%%%%%%%%%%%%%%%%%%%%%%%%%%%%%%%%%%%%%%%%%%%%%%%%
\begin{figure}[t!]
\centerline{\includegraphics[width=0.95\linewidth]{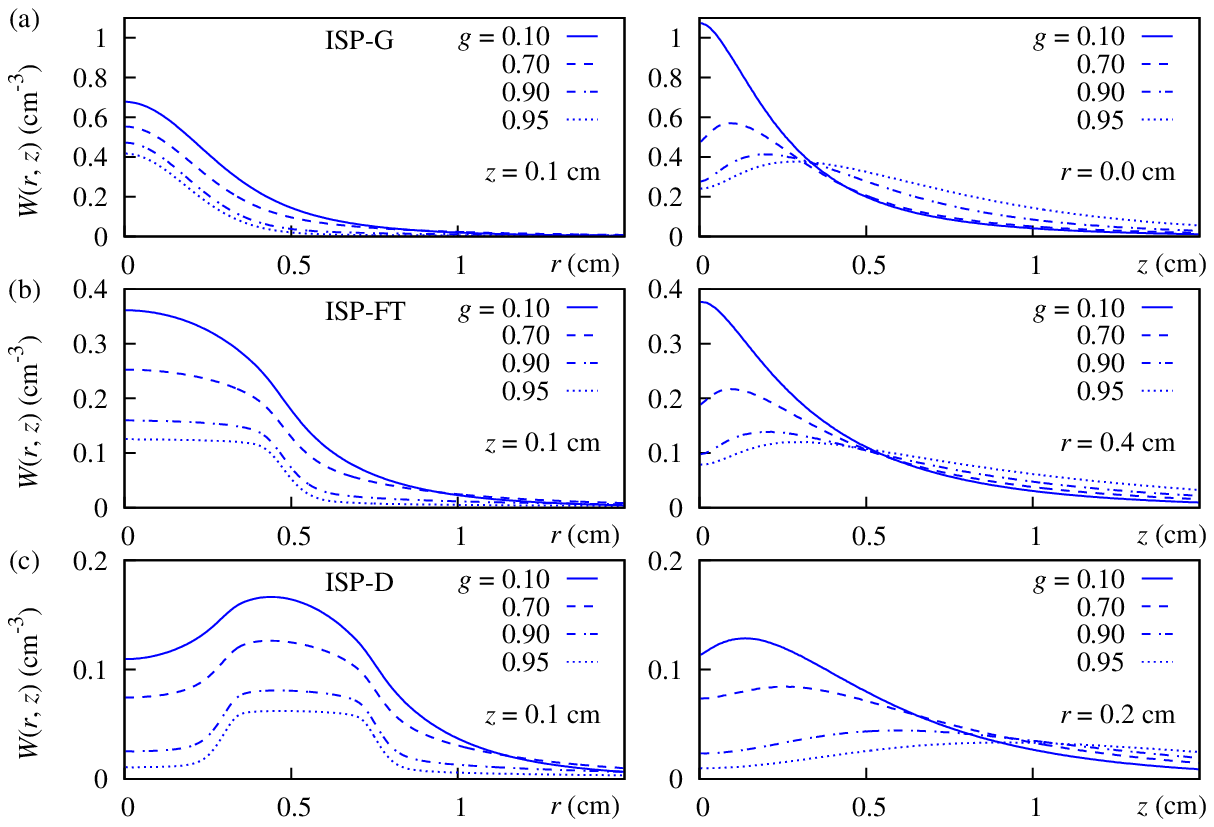} } 
\caption{(Color online) Absorbed energy density $W(r,z)$ at fixed $z$- and 
$r$-slices for the three ISPs used in section \ref{sec:polarConv}.
(a) Gaussian ISP at $z=0.1\,{\rm{cm}}$ (left) and 
$r=0.0\,{\rm{cm}}$ (right) considering different values of the anisotropy $g$,
(b) Flat-top ISP at $z=0.1\,{\rm{cm}}$ (left) and 
$r=0.4\,{\rm{cm}}$ (right),
(c) Donut ISP at $z=0.1\,{\rm{cm}}$ (left) and 
$r=0.2\,{\rm{cm}}$ (right).
}
\label{fig:WSlices}
\end{figure}
%%% END: FIGURE %%%%%%%%%%%%%%%%%%%%%%%%%%%%%%%%%%%%%%%%%%%%%%%%%%%%%%%%%%%%%%%

In Fig.~\ref{fig:beamShapes} we illustrate the Fisk-Johnson convolution
procedure for various anisotropy parameters and three beam shapes:
(i) a Gaussian ISP (G), i.e.\ a special case of Eq.~(\ref{eq:donut}) with
$r_0=0$, $r_1=0$ and $a_1=0.25$ where we used the dFBT parameters $T=4.0$ 
and $N=40$,
(ii) a flat-Top ISP (FT), a special 
case of Eq.~(\ref{eq:donut}) with $r_0=0$, $r_1=0.4$ and $a_1=0.1$ using 
$T=4.0$ and $N=80$,
and,
(iii) a donut (D) ISP with parameters $r_0=0.25$, $r_1=0.6$ and $a_0=a_1=0.05$ 
using $T=4.0$ and $N=150$.
Based on the parameter studies for the forward and reverse dFBT reported in
section \ref{sec:examples}, and by monitoring the rms error for the forward and
immediate backtransformation of the beam profile, yielding $\epsilon_{\rm
rms}<10^{-6}$ (ISP-G), $\epsilon_{\rm rms}=0.003$ (ISP-FT), and, $\epsilon_{\rm
rms}=0.008$ (ISP-D), we opted for the truncation threshold $T$ and summation
truncation parameters $N$ listed above. 
To clarify the behavior of $W(r,z)$ and to illustrate the decrease of $W_{\rm
max}$ as function of $g$, samples of the absorbed energy density at fixed $z$-
and $r$-slices are shown in Fig.~\ref{fig:WSlices}. As one might intuitively
expect, Figs.~\ref{fig:beamShapes} and \ref{fig:WSlices} reveal two tendencies:
(i) for increasing anisotropy $g$, the smoothing of $W(r,z)$ due to scattering
reduces and its absolute values decreases since backscattering is suppressed, 
and, (ii) for increasing $g$, the maximum $W_{\rm max}$ shifts towards deeper 
values of $z$ since scattering is focused on the forward direction.
  A thorough discussion of the characteristics of extended beam profiles and
their use in tissue optics and optoacoustic signal prediction for multilayered
tissues will be presented elsewhere \cite{MelchertOA:2016}.

% 1}}}

% SUMMARY AND CONCLUSIONS 1{{{
\section{Summary and conclusions}
\label{sec:fin}

%%% BEGIN: FIGURE %%%%%%%%%%%%%%%%%%%%%%%%%%%%%%%%%%%%%%%%%%%%%%%%%%%%%%%%%%%%%
%
% TIME EFFICIENCY
%
%%%%%%%%%%%%%%%%%%%%%%%%%%%%%%%%%%%%%%%%%%%%%%%%%%%%%%%%%%%%%%%%%%%%%%%%%%%%%%%
\begin{figure}[t!]
\centerline{\includegraphics[width=0.75\linewidth]{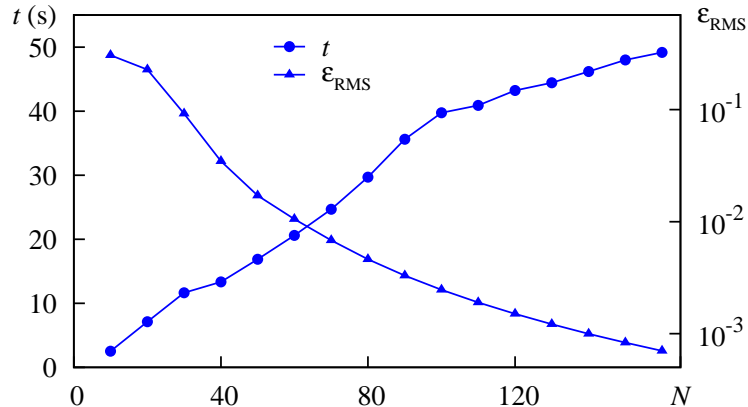} } 
\caption{(Color online) Accuracy and computational efficiency of the Fisk-Johnson
polar convolution as function of the summation truncation parameter $N$ at
fixed $T=4.0$ for the flat-top ISP. The reconstruction RMS error of the ISP
decreases below $10^{-2}$ at approximately $N=50$. At this point, the 
convolution procedure terminates after $t(50)\approx 17\,{\rm{s}}$. For 
comparison: the Cree-Bones procedure used for benchmarking terminates after
time $t_{\rm{CB}}\approx 326\,{\rm{s}}$, highlighting the performance of 
the Fisk-Johnson algorithm.
}
\label{fig:timeEfficiency}
\end{figure}
%%% END: FIGURE %%%%%%%%%%%%%%%%%%%%%%%%%%%%%%%%%%%%%%%%%%%%%%%%%%%%%%%%%%%%%%%

In the presented article we discussed the Fisk-Johnson procedure for computing
a $2$D polar convolution of two radially symmetric functions, based on
efficient discrete approximations to the forward and reverse Fourier-Bessel
integral transform. 
We assessed the efficiency and accuracy of the forward transform, reverse
transform and polar convolution on a set of test functions and applied the
method to a problem from computational biophotonics. Therein, the aim was to
convolve the Green's function material response to an infinitely thin laser
beam to an extended beam profile.
From a point of view of computational efficiency, the presented procedure 
resides between the highly efficient but ISP-restricted direct convolution
(implemented in terms of the {\tt CONV} code \cite{CONV:1997}) and the 
inefficient but accurate straight forward numerical quadrature used for 
benchmarking in section \ref{sec:examples}.
Bear in mind that (time) efficiency is an issue: so as to complete the
convolution procedure for, say, the sampled source volume at $g=0.95$, an
individual convolution has to be carried out for a sequence of $N_z=1414$
consecutive values of $z$, each involving a number of $N_r=1000$ sample points
$r$, see Tab.~\ref{tab:simPars}. For the exemplary case of the previous
flat-top beam profile, Fig.~\ref{fig:timeEfficiency} reveals that the
completion time of the Fisk-Johnson convolution procedure is linear in $M$ with
$t(N)\approx 0.34(1) N \,{\rm{s}}$. In particular, the reconstruction
error of the ISP decreases below $10^{-2}$ at approximately $N=50$. At this
value of $N$, the Fisk-Johnson procedure terminates after $\approx
17\,{\rm{s}}$.  In contrast, note that the Cree-Bones procedure used for
benchmarking in section~\ref{sec:examples} terminates after $\approx
326\,{\rm{s}}$, highlighting the efficiency of the Fisk-Johnson polar
convolution for the considered application. 

Albeit the scientific literature frequently features new algorithms to compute
the above (and further related) transforms for particular scientific
applications, their thorough exploration and implementation in terms of, say,
symbolic computer algebra is rather recent \cite{Dovlo:2015}.  Since the
discrete Fourier-Bessel transform and the polar convolution are valuable
computational tools for the solution of many physical problems with axial
symmetry, and so as to follow the ideal of guaranteeing reproducible results
in scientific publications \cite{Sandve:2013,Barnes:2010}, we considered it
useful to make the research-code for the presented study, along with all
scripts needed to reproduce all figures, publicly available on one of the
authors {\tt gitHub} profile \cite{MelchertGitHub_dFBT:2016}.

% 1}}}

\ack{This research work received funding from the VolkswagenStiftung within the
``Nieders\"achsisches Vorab'' program in the framework of the project ``Hybrid
Numerical Optics''  (HYMNOS; Grant ZN 3061). The software was developed and 
tested under OS X Yosemite (Version: 10.10.3) on a MacBook Air featuring a
1.7GHz Itel Core i5 processor and 4 GB DDR3.}

\section*{References}
\bibliography{commentsBibfile_FourierBesselTrafo,masterBibfile_FourierBesselTrafo}

\end{document}